\documentclass{PoS}

\usepackage{amsmath}
\newcommand{\ba}{\begin{eqnarray}}
\newcommand{\ea}{\end{eqnarray}}

\newcommand{\nn}{\nonumber}

\newcommand{\bs}{\boldsymbol}
\newcommand{\es}{&=&}

\title{Wigner Distributions of Quarks and Gluons}

\ShortTitle{Wigner Distributions}

\author{Jai More\\
       Department of Physics, Indian Institute of Technology Bombay, Mumbai
400076, India \\
        E-mail: \email{more.physics@gmail.com}}

\author{\speaker{Asmita Mukherjee}\\
        Department of Physics, Indian Institute of Technology Bombay, Mumbai
400076, India\\
        E-mail: \email{asmita@phy.iitb.ac.in}}

\author{Sreeraj Nair\\
       Department of Physics, Indian Institute of Science Education and
Research Bhopal\\
Bhopal Bypass Road, Bhauri, Bhopal 462066 India\\
        E-mail: \email{sreeraj.phy@gmail.com}}

\abstract{We present a recent calculation of Wigner distributions of  quarks
and gluons for a quark target state
dressed with a gluon, using  overlaps of light-front wave functions.}

\FullConference{XXV International Workshop on Deep-Inelastic Scattering and Related Subjects\\
		3-7 April 2017\\
		University of Birmingham, UK}

\begin{document}
\section{Introduction}

In quantum mechanics, Wigner distributions \cite{Wigner32} can be interpreted
as semi-classical objects giving a combined position and momentum space
description of a system. Wigner distributions of quarks and gluons 
\cite{Ji03,Lorce11}  in a
nucleon are of interest recently as they are related to generalized
transverse momentum dependent parton distributions(GTMDs) that give the most
general information about the momentum and spin correlations of quarks and
gluons. There is also the possibility of accessing the quark and gluon
orbital angular momentum through Wigner distributions or GTMDs
\cite{Lorce11,hatta12,Hagler03}.  If one integrates
the Wigner distributions over transverse momentum, one obtains impact
parameter dependent parton distributions, which are Fourier transform of
generalized parton distributions (GPDs); integration over transverse
position relates them on the other hand to transverse momentum dependent
parton distributions (TMDs). Quark Wigner distributions have been
investigated in several model calculations
\cite{Lorce11,Liu15,dipankar1,dipankar2}. 
Most of the models do not have a gluonic degree of freedom, and thus 
gluon Wigner distributions are less
explored apart from the small $x$ region. Gluon Wigner distributions, or the
unintegrated gluon correlators have been introduced and discussed in the
context of small $x$ physics in \cite{martin} for diffractive vector meson
production, and in \cite{khoze} for Higgs production at Tevatron and LHC.
Operator definitions of gluon GTMDs for a spin $1/2$ target are given in
\cite{Lorce13}. Gluon GTMDs and Wigner distributions have been studied  
at small $x$ in deeply virtual Compton scattering (DVCS) \cite{Hatta17} and
in hard diffractive dijet production in lepton-hadron scattering
\cite{Hatta16}. In \cite{Hagi} gluon Wigner and Husimi distributions are
discussed. Husimi distributions are positive definite, unlike Wigner
distributions, however, upon integration over $b_\perp$ they do not reduce to
TMDs. Recently, we have  investigated
the quark and gluon Wigner distributions taking the target to be a  quark
dressed with a gluon \cite{Asmita14,Asmita15}. 
This may be thought of as a simple spin $1/2$ composite
state having a gluonic degree of freedom. As a continuation of our project,
we have improved the numerical convergence of the result to remove
an artificial cutoff dependence, as well as extended to transverse
polarization. Here, we report on some of our recent results \cite{More1,
More2}. 

\section{Wigner Distributions in a dressed quark model}\label{WD}

The Wigner distributions of quarks are defined as the Fourier transform of
the quark-quark correlators :
\cite{Meissner09, Lorce11}
\ba
\rho^{[\Gamma]} ({\bs b}_{\perp},{\bs k}_{\perp},x,s,s') = \int \frac{d^2 
\Delta_{\perp}}{(2\pi)^2} e^{-i {\bs \Delta}_{\perp}.{\bs b}_{\perp}}
W_{s\, s'}^{[\Gamma]} ({\bs \Delta}_{\perp},{\bs k}_{\perp},x)
\ea
where ${\bs b_\perp}$ is the impact parameter   conjugate to ${\bs
\Delta}_\perp$, which is the  momentum transfer of the 
state in the transverse direction.
The quark-quark correlator $W^{[\Gamma]}$ for the GTMDs are defined at a
fixed light-front time as
\ba
W_{s\,s'}^{[\Gamma]} ({\bs \Delta}_{\perp},{\bs k}_{\perp},x)
&=&\int \!\!\frac{dz^{-}d^{2} {\bs z}_{\perp}}{2(2\pi)^3}e^{i k.z}
\Big{\langle}p^{+},\frac{{\bs \Delta}_{\perp}}{2},s' \Big{|}
\overline{\psi}(-\frac{z}{2})\Omega \Gamma \psi(\frac{z}{2}) \Big{|}
p^{+},-\frac{{\bs \Delta}_{\perp}}{2},s\Big{\rangle }  \Big{|}_{z^{+}=0}
\ea

The average four-momentum 
of the target state is given by  $P={1 \over 2} (p'+p)$, 
the momentum transfer $\Delta=p'-p$ in the transverse direction. $s$($s'$) is the
helicity of the initial 
(final) target state. The average four momentum of the quark is $k$,
with $k^+=x P^+$, where $x$ is the longitudinal momentum fraction of the
parton. $\Omega$ is the gauge link for color gauge invariance. 
We use light-cone gauge and take the
gauge link to be  unity.

The Wigner distribution of the gluon can be defined as
\cite{Meissner09, Lorce11}
\ba
x W_{\sigma, \sigma'}(x,{k}_{\perp},{b}_{\perp})\es\int \frac{d^2
{\Delta}_{\perp}}
{(2\pi)^2}e^{-i{\Delta}_{\perp}.{b}_{\perp}} \int
\frac{dz^{-}d^{2} z_{\perp}}{2(2\pi)^3 p^+}e^{i k.z} \nn \\
 &\times&\Big{\langle } p^{+},-\frac{{\Delta}_{\perp}}{2},\sigma' \Big{|}
\Gamma^{ij} F^{+i}\Big( -\frac{z}{2}\Big) F^{+j}\Big( \frac{z}{2}\Big)
\Big{|}
p^{+},\frac{{\Delta}_{\perp}}{2},\sigma  \Big{\rangle }  \Big{|}_{z^{+}=0}
\label{eq1}
\ea
\ba
F^{+i} = \partial^+ A^i -  \partial^i  A^+ + g f^{abc} A^+ A^i
\ea
$\Gamma^{ij}=\delta^{ij}$ for unpolarized distribution.  Wigner distribution
for gluons need two gauge links for gauge invariance. As in the quark case,
we use light-cone gauge and take the gauge links to be unity. $\sigma$ and
$\sigma'$ are the helicities of the target state, for unpolarized
distribution $W_{UU}$, we average over the helicity of the target.  Instead
of a proton target, we take the state to be a quark dressed with a gluon. 

A dressed quark state can be expanded in Fock space as 
\ba
  \Big{| }p^{+}, {\bs p}_{\perp}, s \Big{\rangle} &=& \Phi^{s}(p)
b^{\dagger}_{s}(p) | 0 \rangle +
 \sum_{s_1 s_2} \int \frac{dp_1^{+}d^{2}p_1^{\perp}}{ \sqrt{16 \pi^3
p_1^{+}}}
 \int \frac{dp_2^{+}d^{2}p_2^{\perp}}{ \sqrt{16 \pi^3 p_2^{+}}} \sqrt{16
\pi^3 p^{+}}
 \delta^3(p-p_1-p_2) \nn \\[1.5ex]
 &&\times\Phi^{s}_{s_1 s_2}(p;p_1,p_2) b^{\dagger}_{s_1}(p_1)
 a^{\dagger}_{s_2}(p_2)  | 0 \rangle
 \ea
The two-particle light-front wave function LFWF can be written in terms of
 the boost-invariant LFWF as
\ba\sqrt{P^+}\Phi(p; p_1, p_2) = \Psi(x_{i},{\bs q}_{i}^{\perp})\ea
The two-particle LFWF can be calculated in light-front Hamiltonian
perturbation theory.  The Wigner distributions for the dressed quark state
for different combinations of polarizations of the target and the probed
quark are expressed in terms of overlaps of the light-front wave functions.
The single particle sector of the Fock space component plays an important
role when $x=1$. Out of the $16$ leading twist Wigner distributions that
one can define, in our model we obtain only $8$ independent non-zero
functions. They are calculated analytically using LFWFs and the Fourier
transform is done numerically.  Unpolarized gluon Wigner is calculated in the
same approach. Here we present some of our numerical results.

\section{Numerical Results}

In all plots we have used Levin's method to do the numerical integration,
which is a reliable method for oscillatory integrand. We have used the upper
limit of the $\Delta_\perp$ integration to be $\Delta_{max} =
20~\mathrm{GeV}$, and the results are independent of this cutoff.
We choose the mass of the quark to be $m=0.33 ~GeV$.  In all plots we have
integrated over $x$. 

Fig. 1(a) shows the plot of $\rho_{UU}$ in $b$ space. This gives
the distribution of an unpolarized quark in an unpolarized target state. 
Fig. 1(b) shows the plot of $\rho_{UL}$ in $b$ space for a fixed value of
the $k_\perp$. In this model, $\rho_{UL}=\rho_{LU}$. 
This Wigner distribution is related to the orbital angular momentum of the 
quark. A dipole structure is seen similar to chiral quark soliton model and constituent
quark model \cite{Lorce11}. In Fig. 2(a) we have plotted $\rho^x_{UT}$, which gives the
distribution of transversely polarized quark in an unpolarized target;
quark polarization in the $x$ direction. Here we also observe a dipole
structure. TMD limit of $\rho_{UT}$ is the Boer-Mulders function. As we have
taken the gauge link to be unity, we cannot calculate the T-odd
distributions, and Boer-Mulders function is zero in our model.
The behaviour in $b$ space is similar to spectator model \cite{Liu15}.         
In Fig. 2(b) we have shown the gluon Wigner distribution $W_{UU}$, which is
the distribution of an unpolarized gluon in an unpolarized dressed quark. 
As mentioned above, each gluon Wigner distribution needs two
gauge links for gauge invariance. Depending on whether it is a $++$ or $+-$
gauge link combination, it is called a Weizsacker-Williams type or dipole
type gluon distribution. In \cite{Hatta16} it was shown that both of them
give the same orbital angular momentum distribution of the gluon. As we
mentioned, in our
model, we have taken the gauge links to be unity in light-cone gauge. The
gluon distribution has a positive peak at the center of the $b$ space.

\begin{figure}[h]
 \centering 
(a)\includegraphics[width=6cm,height=4cm]{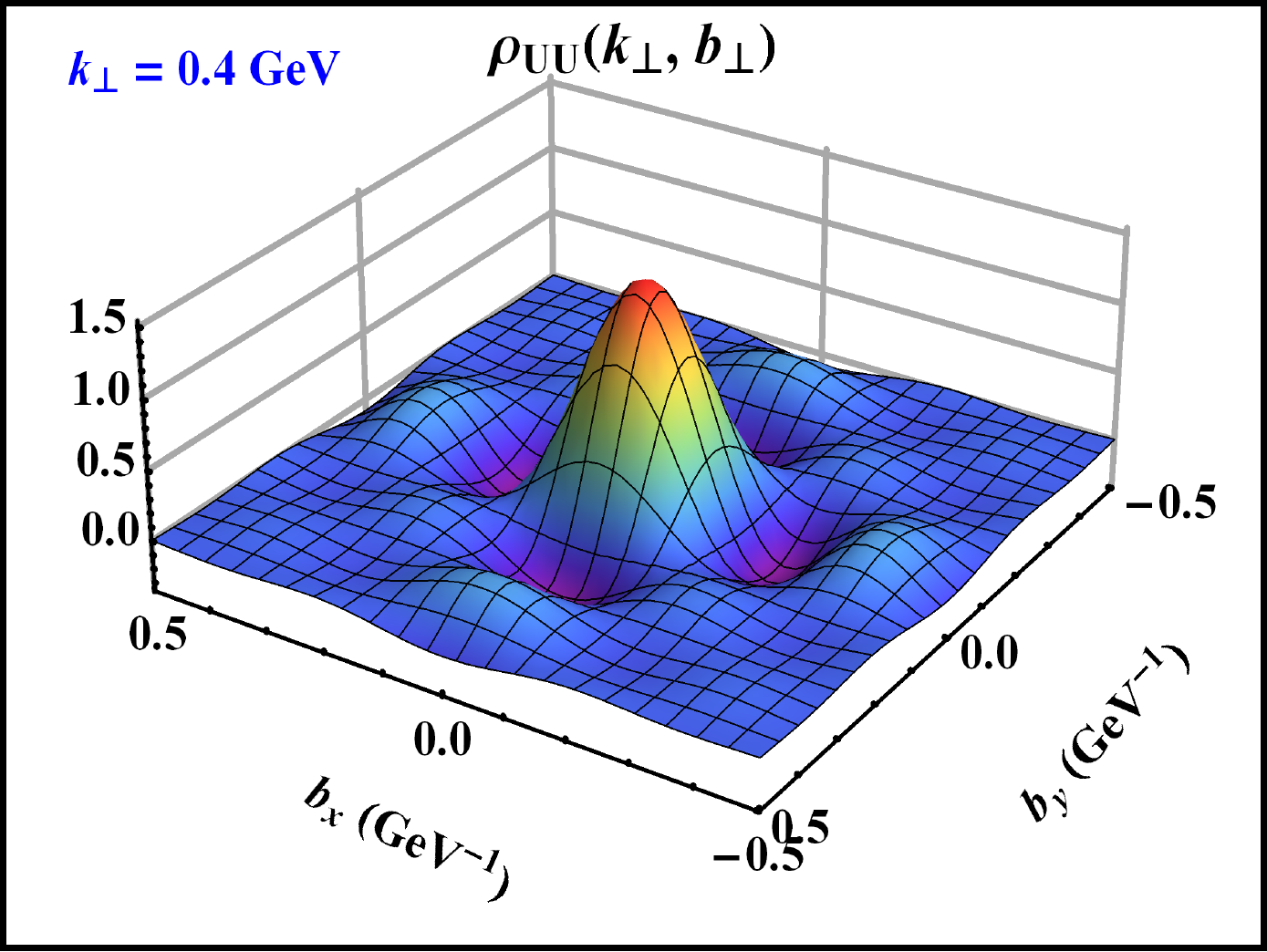}\hskip 0.3cm(b)
\includegraphics[width=6cm,height=4cm]{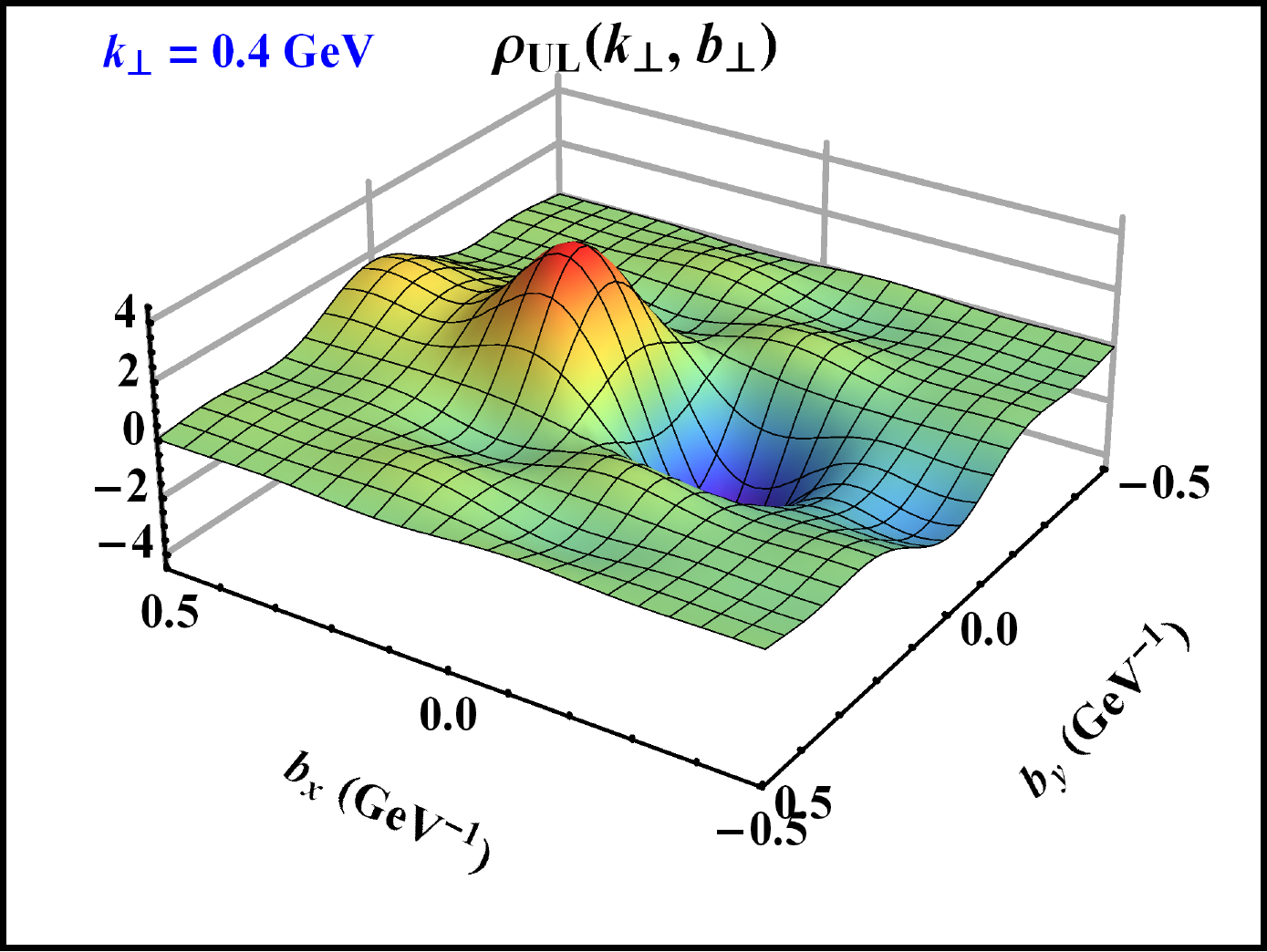}\\[5ex]
 \caption{3D plots of  Wigner distributions $\rho_{UU}({\bs k}_\perp, {\bs b}_\perp)$ and
  $\rho_{UL}({\bs k}_\perp, {\bs b}_\perp)$ at $\Delta_{max} =
20~\mathrm{GeV}$.}
\end{figure} 

\begin{figure}[h]
 \centering
(a)\includegraphics[width=6cm,height=4cm]{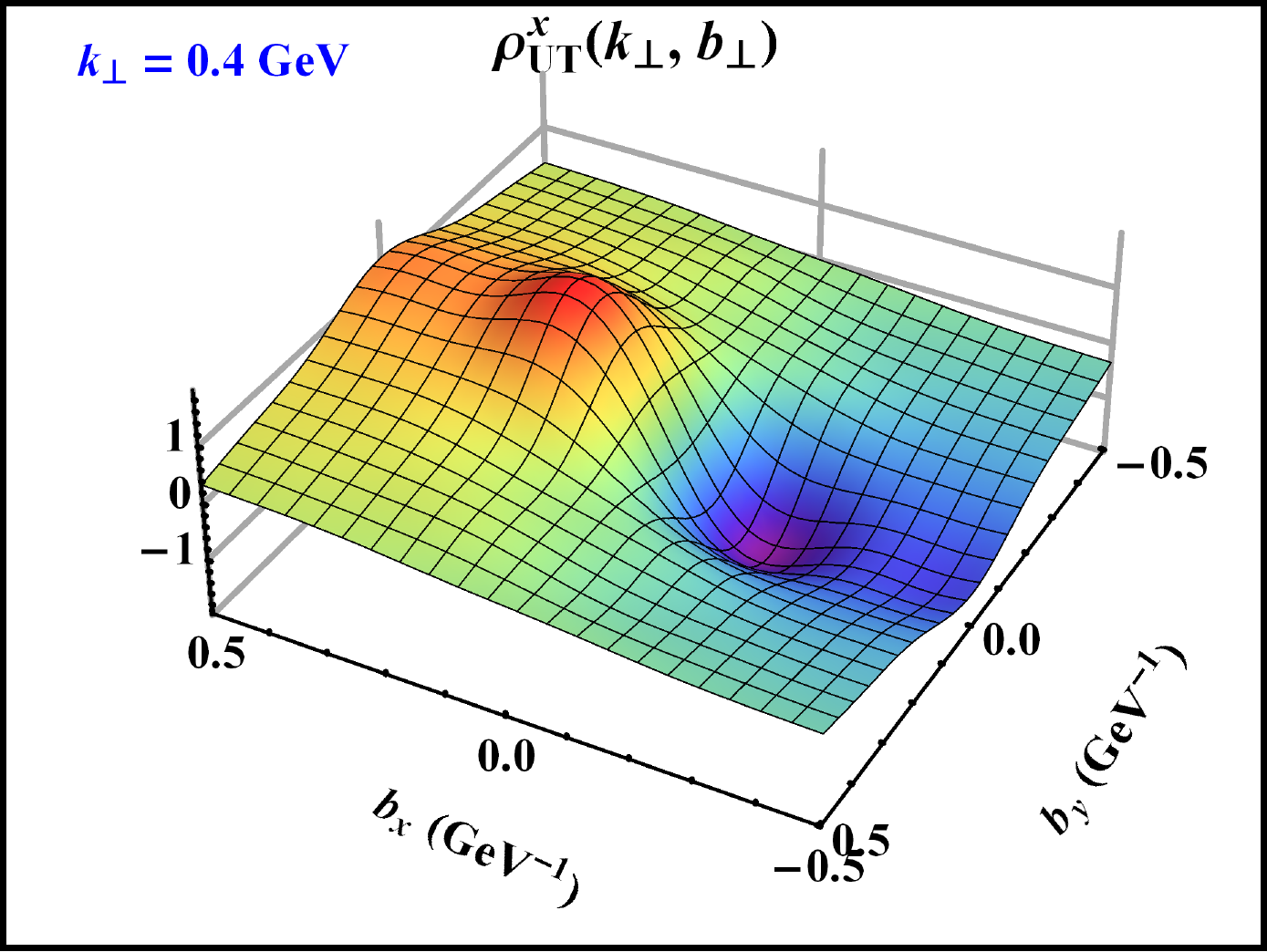}\hskip 0.3cm(b)
\includegraphics[width=6cm,height=4cm]{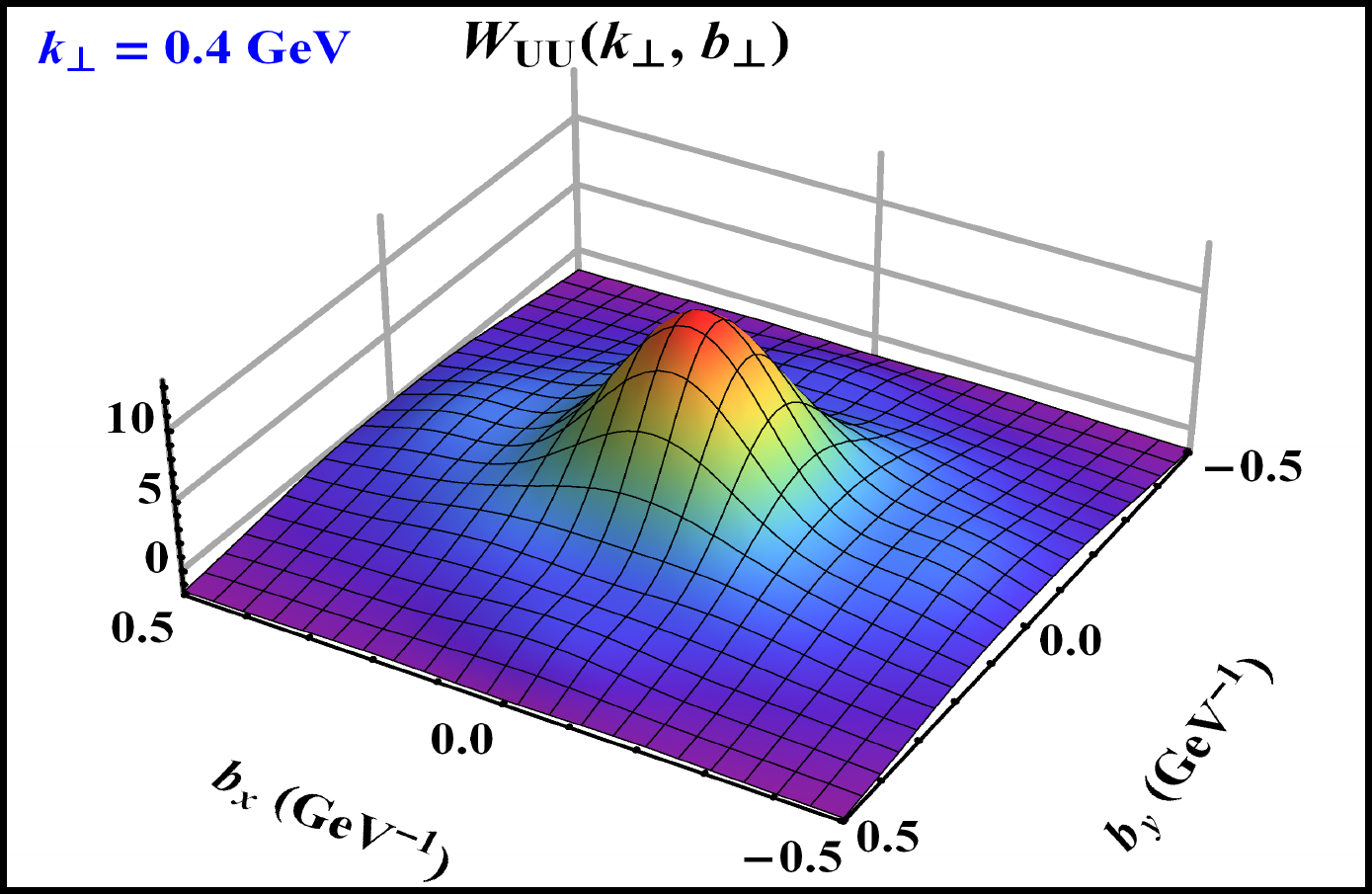}\\[5ex]
 \caption{3D plots of  Wigner distributions $\rho^x_{UT}({\bs k}_\perp, {\bs
b}_\perp)$ and gluon Wigner distribution
  $W_{UU}({\bs k}_\perp, {\bs b}_\perp)$ at $\Delta_{max} =
20~\mathrm{GeV}$.}
\end{figure} 

\section{Conclusion}

We presented a recent calculation of the Wigner distributions of quarks and
gluons in  a dressed quark target model. This represents a composite spin
$1/2$ state with a gluonic degree of freedom. The Wigner distributions are 
expressed in terms of overlaps of LFWFs and using their analytic form one
can calculate them. Better convergence of the results is obtained using
Levin's method of integration, that removes the dependence on an artificial
cutoff on $\Delta_{\perp}$. We presented the Wigner distributions for
unpolarized, longitudinally polarized and transversely polarized quark in an
unpolarized target state. We also presented the unpolarized gluon Wigner
distributions. 

\section{Acknowledgement}

AM would like to thank the organizers of    
"XXV International Workshop on Deep-Inelastic Scattering and Related
Subjects" at the University of Birmingham, UK  for invitation.

\end{document}